\RequirePackage{fix-cm}

\documentclass[11pt]{article}
\usepackage[affil-it]{authblk}
\usepackage{amssymb} 
\usepackage{amsfonts}
\usepackage{geometry}
\usepackage{graphicx}
\usepackage{lipsum}
\usepackage{color}
\usepackage[english]{babel}
\usepackage[utf8]{inputenc}
\usepackage[colorinlistoftodos, color=green!40, prependcaption]{todonotes}%
\usepackage{amsmath}
\usepackage{xcolor,soul}
\usepackage{graphicx}

\usepackage{ulem,soul}
\usepackage[T1]{fontenc}
\usepackage{multirow}

\usepackage{soul}
\RequirePackage{graphicx}
\RequirePackage{color}
\usepackage[pdftex, pdftitle={Article}, pdfauthor={Author}]{hyperref} 
\begin{document}

\title{
On the Gaussian Assumption in the Estimation of Parameters for Dark Energy Models
}

\author{Fabiola Arevalo}
\affil{Universidad Mayor, Temuco, Chile.\thanks{Electronic address: \texttt{fabiola.arevalo@umayor.cl}}}

\author{Luis Firinguetti}
\affil{Universidad del Bio-Bio, Concepción, Chile.\thanks{Electronic address: \texttt{lfiringu@ubiobio.cl }}}

\author{Marcos Peña}
\affil{Universidad de Magallanes, Punta Arenas, Chile.\thanks{Electronic address: \texttt{marcos.pena@umag.cl }}}





\date{Received: date / Accepted: date}

\maketitle

\begin{abstract}
Type Ia supernovae have provided fundamental observational data in the discovery of the late acceleration of the expansion of the Universe in cosmology. However, this analysis has relied on the assumption of a Gaussian distribution for the data, a hypothesis that can be challenged with the increasing volume and precision of available supernova data. 
In this work, we rigorously assess this Gaussianity hypothesis and analyze its impact on parameter estimation for dark energy cosmological models. We utilize the Pantheon+ dataset and perform a comprehensive statistical, analysis including the Lilliefors and Jarque-Bera tests, 
{to assess the normality of both the data and model residuals.}

We find that the Gaussianity assumption is untenable and that the redshift distribution is more accurately described by a t-distribution, as indicated by the Kolmogorov Smirnov test.  {Parameters are estimated for a model incorporating a nonlinear cosmological interaction for the dark sector. The free parameters are estimated using multiple methods, and bootstrap confidence intervals are constructed for them.}

\end{abstract}

\section{Introduction}
\label{intro}
Cosmology is currently in an era of unprecedented precision measurements. Almost a century ago the expansion of the Universe was observationally confirmed and over two decades ago the accelerating expansion was announced,  based on Type Ia supernova data \cite{riess1998} and \cite{perlmutter1999}.
 The standard model of cosmology, based on General Relativity, $\Lambda$CDM, considers a nearly flat expanding Universe that at the time is composed of approximately 73\% component of dark energy (DE) and another 23\% component of dark matter (DM), these components are referred to as the Dark Sector \cite{Tsujikawa:2010sc} which drives the current accelerated expansion of the Universe. The unknown nature of these dark components represents one of the challenges in modern cosmology and constitutes the focus of this work from a statistical perspective.
  
The Gaussian assumptions underlying these data,  particularly in the search for the best fit parameters, warrants a more detailed examination given the increased volume of available data.


According to \cite{Scolnic:2021amr}, the Pantheon+ dataset, includes an increased number of low-redshift supernovae compared to previous datasets.
This increased data volume highlights a more evident bimodality in the distribution of supernovae with respect to redshift.
 
 The assumption of normality (or gaussianity) in supernovae analysis is a central topic adressed in this work. This extends the work of \cite{Singh:2016xib} on non gaussianity in the error distribution and of \cite{dainotti2024} and \cite{Singh:2025seo} on the gaussianity assumption more broadly on line with current research on the topic \cite{Hussain:2025nqy}.

The $\Lambda$ Cold Dark Matter ($\Lambda$CDM) model is widely considered a simple yet successful model that fits cosmological data remarkably well, although it has some known issues \cite{Bull:2015stt}. 
There is a plethora of modified candidates, known as dark energy models, which are of significant interest. In this work, we focus on cosmological interaction from a phenomenological perspective due to their versatility and availability of analytical solutions  \cite{Wang:2016lxa}. The dark sector interacts only gravitationally, therefore one could expect that modifying its gravitation dynamics could offer a promising avenue for understanding or predicting some of its key features. We will focus on a sign-changeable interaction, extending the work of \cite{Arevalo:2022sne} by performing data analysis without making any specific assumptions about the distribution of supernovae data.


We will establish a sign-changeable interacting dark sector model to describe the evolution of the late Universe. Our aim is to determine the values of cosmological parameters such as the expansion rate of the Universe, the density of matter, and the density of dark energy, among others.  

The subsequent sections of this study will describe the relevant variables, their distributions, and their correlations, followed by the estimation of model parameters and their corresponding confidence intervals.


\section{Cosmological Framework
} \label{S2}
Theoretical cosmology models the Universe using the Einstein Field Equations (EFE) with 
an ideal matter content. For a flat Friedmann-Lemaitre-Robertson-Walker metric this leads to the following set of equations
\begin{equation} \label{ec1}
3H^2=\kappa \rho, \dot H=-\frac{\kappa}{2}(\rho+p), \dot \rho+3H(\rho+p)=0.
\end{equation}
These are the Friedmann equations and the conservation equation respectively, where $H =\dot a/a$ is the Hubble parameter, $a$ is the scale factor and the dot denotes the derivative of $a$ with respect to cosmic time $t$. 

The total energy density $\rho=\sum \rho_i$  and pressure $p=\sum p_i$ depend on the matter content that we are modeling, which in turn determines the Hubble parameter and the overall dynamics of the model.

One of the motivations for studying cosmological interactions is to address the cosmic coincidence problem: Why the energy densities of dark energy $\rho_x$ (DE) and dark matter $\rho_m$ (DM) are of the same order of magnitude today? For a comprehensive review of interactions within the dark sector, see \cite{Wang:2016lxa} and references therein. Various specific choices have yielded a large number of interactions (both linear and nonlinear) with known late-time behavior
A key challenge lies in distinguishing which interactions, or families of interactions, provide a better overall description of the cosmological evolution

In this work, we will model the dark content neglecting radiation and other minor components, as we are interested in late Universe cosmology. Consequently, the total energy density is defined by $\rho=\rho_x+\rho_m$. In this scenario, cosmological interactions are denoted by $Q$ and can be introduced as a phenomenological coupling. 
This coupling is generally defined as a function of system variables, such as the individual or total energy densities $\rho_x$ and $\rho_m$, the Hubble parameter $H$, derivatives of these quantities, or even linear combinations including a constant. We use a dependence on the deceleration parameter, defined as $q=-1-\frac{\dot H}{H^2}$, where a change of sign can be obtained naturally \cite{Arevalo:2022sne}, following the work of \cite{Arevalo:2019axj}.

The energy conservation equation (\ref{ec1}) can be decoupled into two equations by introducing the phenomenological interaction term $Q$ to model the conversion of energy between dark energy and dark matter
\begin{eqnarray}
\dot \rho_m + 3H \rho_m \gamma_m = -3HQ, \label{rhom} \\
\dot \rho_x + 3H \rho_x \gamma_x = 3HQ.\label{rhox}
\end{eqnarray}
Here, $\gamma_i$ is the equation of state parameter, defined as $p_i=(\gamma_i-1) \rho_i$.
When $Q$ is positive, it indicates an energy transfer from dark matter to dark energy. When an interaction is given as an Ansatz, it modifies the resulting Hubble parameter $H$ as a function of the scale factor $a$. {When $Q=0$, there is no cosmological interaction. Additionally, when $\gamma_m=1$, we obtain $\omega CDM$ and when $\gamma_x=0$, we obtain $\Lambda CDM$.} Following \cite{Arevalo:2019axj} and \cite{Chimento:2009hj} we can write equations (\ref{rhom}) or (\ref{rhox}) in terms of a second-order differential equation for the total energy density:  
\begin{equation}
\rho''+\rho'\gamma_m+\gamma_x+\rho \gamma_m \gamma_x=Q(\gamma_m-\gamma_x). \label{rho2}
\end{equation}
Extending the work of \cite{Arevalo:2019axj}, we propose a cosmological interaction function given as
\begin{equation}
    Q=\frac{q}{3H}\left( \alpha \dot \rho_x+3\beta H \rho_x\right)=q\left( \alpha \rho_x'+\beta  \rho_x\right),
\end{equation}
where $\alpha$ and $\beta$ are constants. {If both these constants are null, it implies that there is no cosmological interaction present.} These interactions naturally undergo a sign change at some point during their evolution, due to the $q$ term. By focusing in a particular case for $Q$, $\alpha=\frac{2}{3\gamma_m-2}$, we can obtain a solution for (\ref{rho2}) using the methods described in \cite{Chimento:2009hj}. We then obtain the total energy density as:
\begin{equation}
    \rho=c_1a^{3/4 \lambda_+}+c_2a^{3/4 \lambda_-},
\end{equation}
where $c_1$ and $c_2$ are integration constants {and $\lambda_\pm$ are constants that depend on $\gamma_i$ and $\beta$. Even if the parameter $\beta$ is zero, the total interaction remains non-zero, given our choice of $\alpha  \neq 0$.}

When we consider limits for the interaction and deceleration parameter, it behaves as a decreasing function, the interaction for the scale factor {$a\rightarrow \infty$ tends to zero or $\pm \infty$ depending on the sign of the constants.}

\section{Statistical and observational analysis}
This paper aims to estimate the model parameters utilizing three distinct methods: the restricted gradient method \cite{nesterovintroductory}, the Expectation-Maximization (EM) method \cite{mclachlan2007algorithm} and the grid method. All calculations were performed using R (The R Project for statistical computing: https://www.R-project.org/). We analyze the \textit{Pantheon}+ data set (see \cite{Scolnic:2021amr}). 
We begin by presenting descriptive statistics for the distance modulus and redshift in Table \ref{T1}.
 
\begin{table}[!ht]
    \centering
    \begin{tabular}{|c|c|c|c|c|c|c|c|}
    \hline
    \textbf{Variable} & \textbf{Min. } & \textbf{1st } & \textbf{Median} & 
       \textbf{Mean} & \textbf{3rd } & \textbf{Max.} & \textbf{Standard } \\ 
   & \textbf{Value} & \textbf{Quartile} &  & 
        & \textbf{Quartile} & \textbf{Value} & \textbf{Deviation} \\ 
     \hline
      \textbf{DM} & 33.21 & 38.91 & 40.46 & 40.05 & 41.78 & 46.18 & 2.647864 \\
      \hline
       \textbf{redshift} & 0.011 & 0.13168 & 0.24855 & 0.32231 & 0.42271 & 2.26 & 0.2859663 \\
       \hline
    \end{tabular}
    \caption{Summary statistics for Distance Modulus DM and redshift variables.}
    \label{T1}
\end{table}

\begin{figure}[h!]
    \centering
    \includegraphics[width=0.9\linewidth]{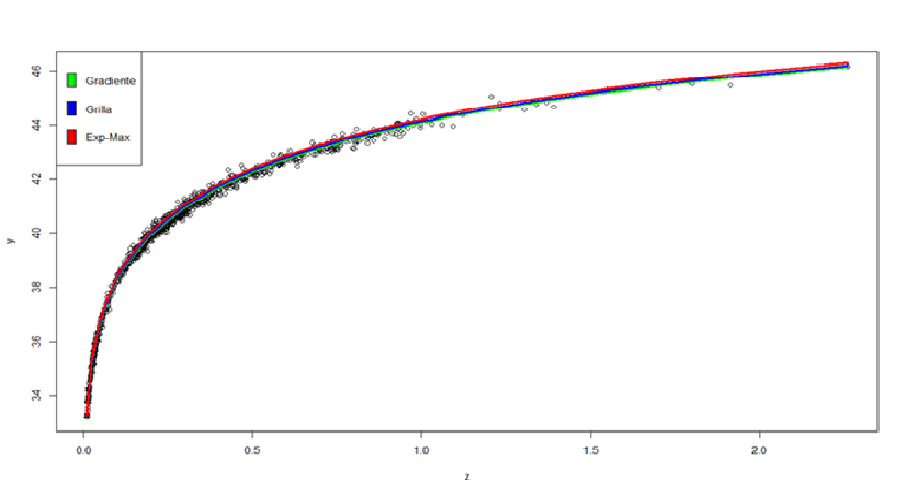}
    \caption{
    Comparison of distance modulus fit using Grid, EM and Gradient methods for Model I
    }
    \label{Figura1}
\end{figure}

We found that the Pearson correlation and the Spearman rank correlation coefficients between these two variables are 0.845 and 0.997, respectively. It is noteworthy that the Spearman rank correlation coefficient is higher than the Pearson correlation. This is attributed to the Spearman coefficient's superior ability to capture monotonic non-linearity compared to the Pearson correlation, as evidenced by the clear non-linear relationship between redshift and distance modulus shown in Figure \ref{Figura1}. For a definition of the Spearman rank correlation coefficient, refer to \cite{spearman}.

Prior to proceeding with the analysis, it is pertinent to note a common, albeit implicit, assumption in cosmological model estimation: that the data follow a normal or Gaussian distribution.

 This assumption often leads to the use of the minimum chi-squared method for parameter estimation and hypothesis testing (see, for instance \cite{verde2010statistical}). However, this assumption does not seem to be correct. In fact Figure 1 in \cite{Scolnic:2021amr}, {suggests that the distribution of the redshift} is neither normal nor bimodal. It may well be a mixture of two normal or perhaps of two t distributions. On this point, see also \cite{dainotti2024}.

The model parameters were estimated using three distinct approaches described earlier.
{In Figure \ref{Figura1}, we show the fitted curves for the three methods.} The parameter estimates according to all three methods for the first interaction model are summarized in Table \ref{Tabla2}. The results of the estimates are presented for the model with same number of free parameters. An * denotes a parameter chosen a priori based on the literature. {It is worth noting that all methods yield fairly consistent parameter estimates, with Models I and V providing the best fits to the data.}

\begin{table}[h!]
\centering
\begin{tabular}{|l|l|l|l|l|l|l|l|}
\hline\noalign{\smallskip}
&\textbf{ Method} & \textbf{$\Omega_m$}& \textbf{$H_0$ }& \textbf{$\gamma_m$} & \textbf{$\gamma_x$ }&\textbf{$\beta$} & GoF\\
\noalign{\smallskip}\hline\noalign{\smallskip}
    & Grid      &0.34    &0.74    &1.12     &-0.28   &0.15* &0.005869828\\
I   & EM        &0.32    &0.74    &1.14     &-0.24   &0.15* &555.3492\\
    & Gradient  &0.3371  &0.7449  &1.1239   &-0.2760 &0.15* &0.005800905\\
\noalign{\smallskip}\hline
    & Grid      &0.34    &0.66    &1*       &-0.11   &0.15* & 0.01711745\\
II  & EM        & 0.35   & 0.68   &1*       &-0.13   &0.15* &552.7129\\
    & Gradient  & 0.3485 & 0.7294 &1*       &-0.1241 &0.15* &0.01752270\\
\noalign{\smallskip}\hline
    & Grid      &0.26    &0.74    &1.19     &-0.151* &0.15* &0.007067298\\
III & EM        & 0.25   & 0.72   &1.22     &-0.151* &0.15* &555.2940\\
    & Gradient  & 0.2576 &0.7451  &1.1940   &-0.151* &0.15* &0.006947955\\
\noalign{\smallskip}\hline
    & Grid      &0.34    &0.66    &1*       &-0.115* &0.15* &0.0172477\\
 IV & EM        &0.35    &0.73    &1*       &-0.115* &0.15* & 552.5443\\
    & Gradient  &0.3482  &0.6626  &1*       &-0.115* &0.15* &0.01653571\\
\noalign{\smallskip}\hline
    & Grid      &0.26    &0.74    &1.2*     &-0.13*  &0.15* &0.006729246\\
 V  & EM        &0.26    &0.68    &1.2*     &-0.13*  &0.15* &555.286\\
    & Gradient  &0.2602  &0.7456  &1.2*     &-0.13*  &0.15* &0.006659599\\
\noalign{\smallskip}\hline
    & Grid      &0.25    &0.67    &1*       &0*      &-0.45 & 0.01460781\\
 VI & EM        &0.25    &0.67    &1*       &0*      &-0.42 & 552.778\\
    & Gradient  &0.2517  &0.6688  &1*       &0*      &-0.4471 & 0.01507553\\
\noalign{\smallskip}\hline
\end{tabular} \label{Tabla2}
\caption{Results from the analysis with different methods for interaction $Q$. Fixed a priori parameters are denoted with *. Goodness of Fit (GoF) refers to the minimum sum of squares for the Grid and Gradient  methods and to the maximum for the EM algorithm.}
\end{table}

\subsection{Residual Analysis}
In Table \ref{Tabla3}, we present some descriptive statistics for the residuals of the model estimated by restricted gradient method. 
\begin{table}[h!]
\centering
\begin{tabular}{|l|l|}
\hline
\textbf{Statistic} & \textbf{Value} \\
\hline
Minimum Value & -0.47209 \\
First Quartile & -0.10585 \\
Median & -0.02651 \\
Mean & -0.02135 \\
Third Quartile & 0.06298 \\
Maximum Value & 0.48433 \\
Standard Deviation & 0.1425124 \\
\hline
\end{tabular}
\caption{Summary Statistics} \label{Tabla3}
\end{table}

\begin{figure} 
    \centering
    \includegraphics[width=0.8\textwidth]{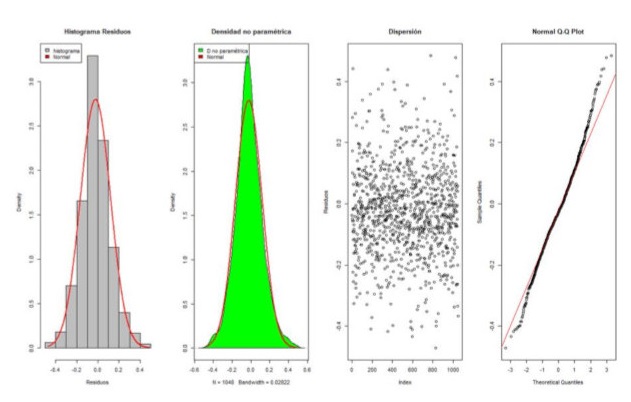}
    \caption{Residual Density}
    \label{fig:2}
\end{figure}

These statistics suggest that the residual distribution may not be symmetrical and is not centered at the origin. Furthermore, a standard deviation of 0.14, provides some evidence that the residuals have a platykurtic distribution, as may be observed from Figure \ref{fig:2}.
The histogram of the residuals and the non-parametric density estimate both reveal discrepancies in kurtosis and tails compared to a superimposed normal distribution.
 
Additionally, the Q-Q plot indicates that the observed residual distribution shows differences in the tails with the normal distribution.

Given that the normality assumption is a common practice in cosmological studies such as JLA \cite{SDSS:2014iwm} and Pantheon \cite{Scolnic:2021amr},
we decided to test this assumption and carried out the Lilliefors \cite{lilliefors} and the Jarque-Bera \cite{jarquebera} tests to check whether the residuals actually follow a normal distribution. The Lilliefors test resulted in a statistic D = 0.039194 with a p-value of 0.0007, leading to the rejection of the normality hypothesis. For the case of the Jarque-Bera test, the statistic resulted in $JB=26.19$ with a $p-value=10^{-16}$, thus, consistent with the previous test, the hypothesis of normality is rejected.

Consequently, we fitted a Skew-t distribution to the residuals. The probability density function of this distribution is given by the following expression (see \cite{davis2015}:	
\begin{equation}
f(x; \mu, \sigma, \lambda, q) = \frac{\Gamma ( \frac{1}{2}+q )}{v \sigma (\pi q)^{\frac{1}{2}} \Gamma (q) \left[ 1 + \frac{\left| x-\mu + m \right| ^2}{q (v \sigma) ^2 (1+\lambda \textup{sign}(x-\mu + m))^2}\right]^{\frac{1}{2}+q}}
\end{equation}
with
\begin{eqnarray*}
&m& = 2\lambda v \sigma \frac{ q^{\frac{1}{2}} \Gamma (q-\frac{1}{2})}{\pi^{\frac{1}{2}} \Gamma (q)}, \\
&v &= \frac{1}{q^{\frac{1}{2}} \sqrt{ (1 + 3 \lambda^2) \frac{1}{2q-2} -\frac{4 \lambda^2}{\pi} \left( \frac{\Gamma ( q - \frac{1}{2} )}{ \Gamma ( q )} \right)^2 }}
\end{eqnarray*}
and where:
\begin{itemize}
		\item $\mu$ is the location parameter.
		\item $\sigma $ is the scale parameter.
		\item $\alpha$ is the skewness parameter.
	\end{itemize}	

{The estimated parameters were:  $\mu=-0.03$, $\sigma=0.12$, $\lambda= 0.04$ and $q=6$ degrees of freedom}. The fitted distribution is presented in the {following figure.}
\begin{figure}[h!]
    \centering
    \includegraphics[width=0.9\linewidth]{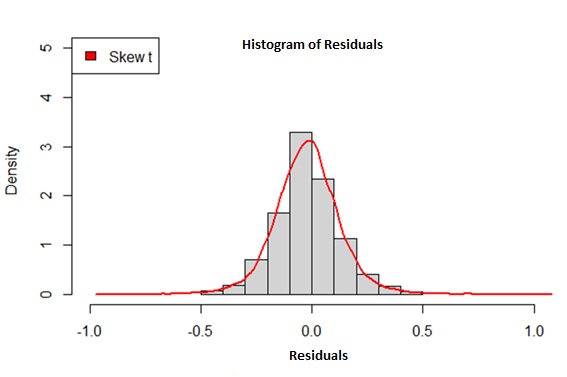}
    \caption{Fitted Skew t Distribution}
    \label{fig:3}
\end{figure}
The Kolmogorov-Smirnov goodness of fit test statistic was 0.018258 with a $p-value =0.9099$. Therefore, the $Skew-t$ hypothesis for the residuals is not rejected. Note that the distribution is essentially symmetric around zero. But it is also true that the distribution has heavier tails than the normal distribution, as can be deduced from the fact that the distribution has only  six degrees of freedom.

To construct confidence intervals for the parameters of the $\Lambda$CDM cosmological model, we applied the bootstrap method. {The procedure was as follows:
First, $B$ bootstrap samples were generated by resampling with replacement from the original Pantheon dataset, which contains 1,358 observations.} Each bootstrap sample consisted of 1,358 data points, allowing for repetitions and the potential omission of some original observations.
For each bootstrap sample $b=1, \ldots, B$, the parameters were estimated by minimizing the sum of squared residuals, resulting in a set of estimated parameters $(\hat{\Omega}_b, \hat{H}_b, \hat{\gamma}_{b,m}, \hat{\gamma}_{b,x})$. Table \ref{Tabla4} presents the average values for each estimated parameter obtained from this process.

\begin{table}[h!]
\centering
\begin{tabular}{|l|l|}
\hline
\textbf{Parameter} & \textbf{ Bootstrap Average} \\
\hline
    $\bar{{\Omega}}$    & 0.3335799 \\
    \hline
    $\bar{H}$         & 0.7399400 \\
    \hline
    $\bar{\gamma}_m$  & 1.1247780\\
    \hline
    $\bar{\gamma}_x$  &-0.2752219 \\
\hline
\end{tabular}
\caption{ Bootstrap Samples Averages} \label{Tabla4}
\end{table}

{From the empirical distributions of the estimators derived from the bootstrap samples, the 95\% confidence intervals  for each parameter were calculated using the percentile method. The intervals obtained are presented in Table \ref{Tabla5}.}

\begin{table}[h!]
\centering
\begin{tabular}{|l|l|l|}
\hline
\textbf{Parameter} & \textbf{2.5\%} &\textbf{97.5\%} \\
\hline
    $\hat{\Omega}$    & 0.3234277 & 0.3425161 \\
    \hline
    $\hat{H}$         & 0.7239443 & 0.7489723 \\
    \hline
    $\hat{\gamma}_m$  & 1.1055430  & 1.1367230 \\
    \hline
    $\hat{\gamma}_x$  & -0.2944569 & -0.2632766 \\
\hline
\end{tabular}
\caption{95\% Bootstrap  Confidence Intervals} \label{Tabla5}
\end{table}

\section{Analysis of Results}

\begin{figure*}[!ht]
      \includegraphics[width=0.45\textwidth]{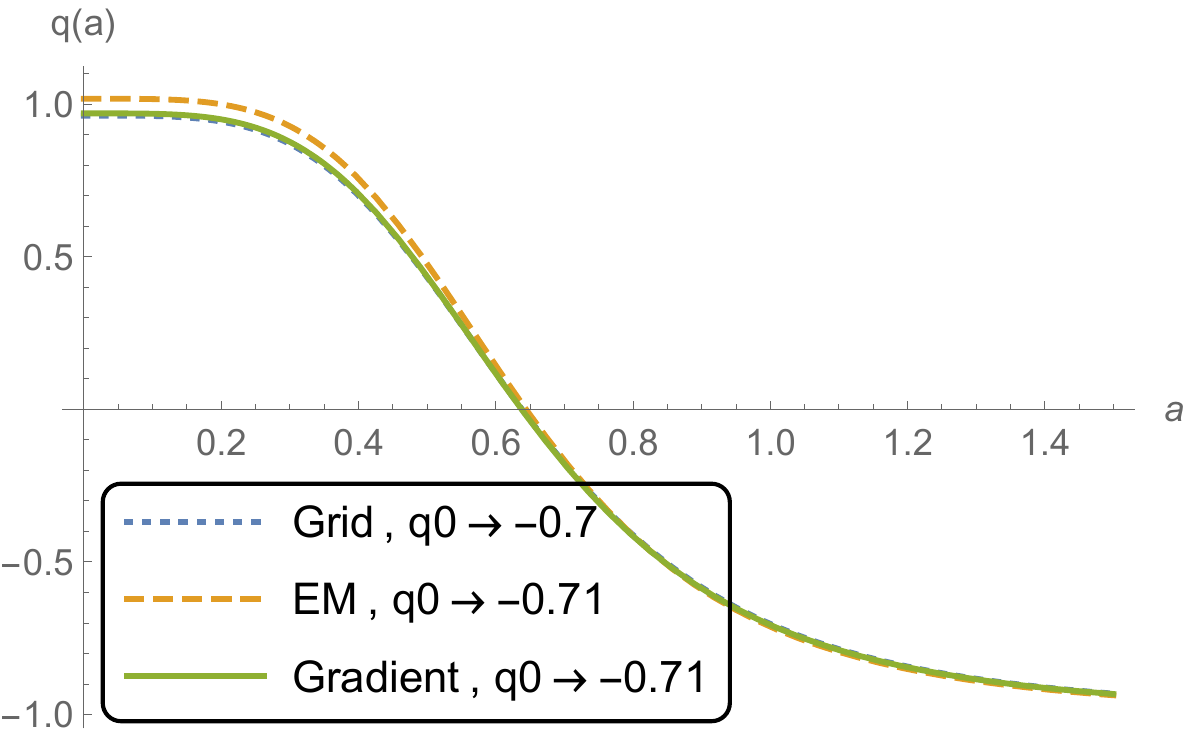}
        \includegraphics[width=0.45\textwidth]{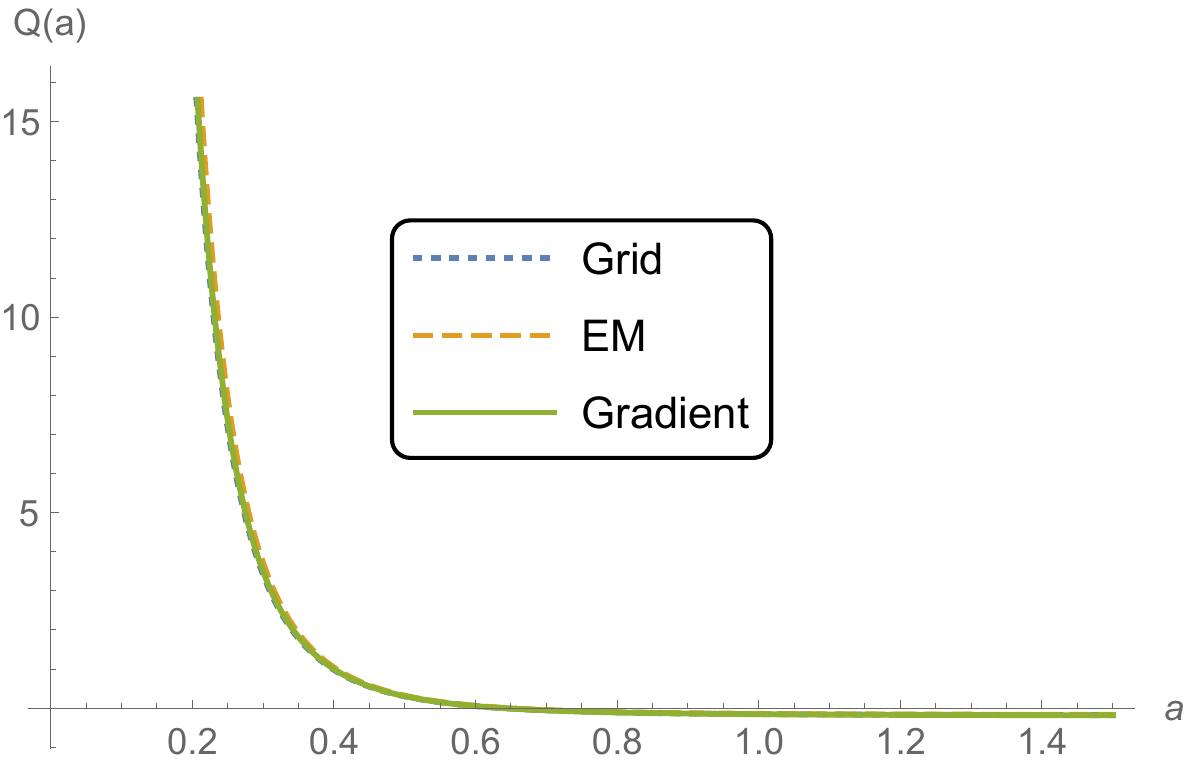}
        \caption{The deceleration parameter and the interaction for interaction 1 are plotted using the results of Table \ref{T1}, case I, 4 free parameters.
}
\label{fig:4}       
\end{figure*}

\begin{figure*}[!ht]
 \includegraphics[width=0.45\textwidth]{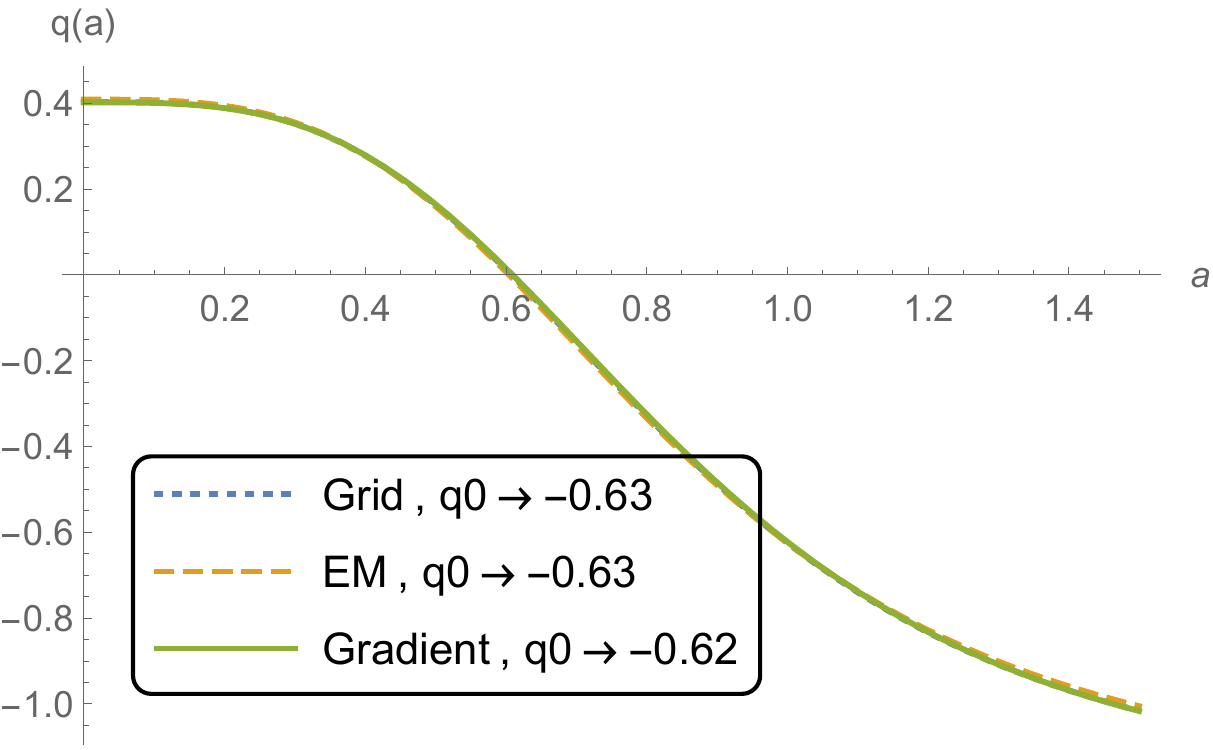}
        \includegraphics[width=0.45\textwidth]{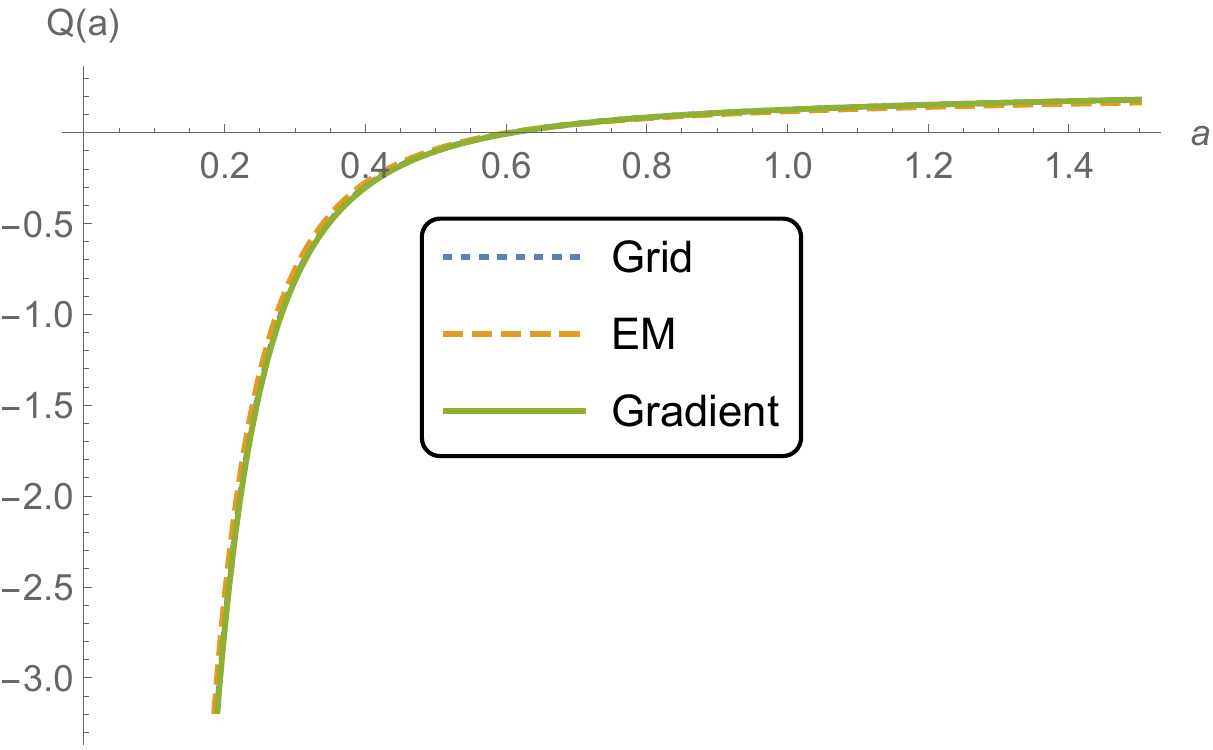}
\caption{The deceleration parameter and the interaction for interaction 1 are plotted using the results of Table \ref{T1}, case VI, 3 free parameters for $\gamma_x=0$ and $\gamma_m=1$.}
\label{fig:5}       
\end{figure*}

Using the solution from section \ref{S2} we obtain an analytical expression for the individual energy densities $\Omega_x=\rho_x/(3H_0^2)$ and $\Omega_m=\rho_m/(3H0^2)$, 
the deceleration parameter $q$ and the cosmological interaction $Q$, all these in terms of the scale factor. We plotted the latter two functions for the minimum in case I for {Figure \ref{fig:4} and case VI for Figure \ref{fig:5}}. We choose these cases as the first corresponds to the best fit and the latter includes the scenario of $\beta$ negative and $\Omega$ positive. We also include in both figures the value of $q_0$, the change of sign of the deceleration parameter. All curves present a similar evolution for different methods. 

We explored other priors  corresponding to cases II to V in Table \ref{Tabla2}. Their behavior is highly similar. We then sought a range where this energy density remains positive, thereby restricting the search to a smaller parameter space. By setting $\gamma_x=0$ and $\gamma_m=1$, we imposed the conditions  $\Omega_m<2/3$ and $\beta<6(\Omega_m-1)\Omega_m/(3\Omega_m-2)$
and searched for a new minimum. {This is case VI, where $\beta<0$. In Figure \ref{fig:5} we fixed the parameters $\gamma_x=0$ and $\gamma_m=1$.} This scenario corresponds to a local minimum, although not the global best fit.

{In all cases, we observe that there is a sign change for the interaction, which is expected given the role of the deceleration parameter in its formulation}. {In Figure \ref{fig:4}} we see that $Q$ is positive in the past, which means that there is energy transfer from dark matter to dark energy, which then switches slightly to a negative interaction, where the transfer of energy is in the opposite direction. On the other hand, we see that {Figure \ref{fig:5}} reveals a reversed scenario: the interaction is initially negative and subsequently becomes positive. This is explained by the result of $\beta$ negative.

\section{Final Remarks}
We studied a model of interacting dark sector with rigorous statistical tools for supernova data. These observational analyses lead to plausible cosmological scenarios that are consistent with previous results. 

The obtained model exhibits late-time acceleration for all priors, presenting a transition from decelerated to accelerated close to -0.7, and remains in an accelerating phase today. The cosmological interaction $Q$ evolves towards zero at late times and tends to a small positive constant, Figure \ref{fig:4}. In one of the scenarios the interaction changes sign, as presented in Figure \ref{fig:5}. 


{Each model offers a plausible mechanism to alleviate the cosmic coincidence problem, as evidenced by the stability and late behaviour of the coincidence parameter.}

The nonlinearity of $\Lambda$CDM model poses certain challenges to its estimation. The statistical analysis reveals that the Gaussian assumption, common in the empirical study of cosmological models is untenable. We have found that a better distribution to fit the data is a slightly skew t distribution. Notably, this distribution consistently exhibits heavier tails than a normal distribution. Therefore, a more appropriate approach for conducting an inferential study requires the use of nonparametric methods such as the Spearman correlation coefficient, the Kolmogorov-Smirnov test, and the construction of bootstrap confidence interval. 
These methods do not require the assumption of a particular underlying distribution. Consequently, we have been able to derive precise and cosmologically consistent parameter estimates.

\label{fig:1}       
%
%




\end{document}